\documentclass[12pt]{article}
\usepackage{graphicx}

\newcommand\pubnumber{DPF2015-182}
\newcommand\pubdate{\today}

\def\iowa{Department of Physics and Astronomy, 
University of Iowa \\ Iowa City, IA 52242, USA}

\def\Title#1{\begin{center} {\Large #1 } \end{center}}
\def\Author#1{\begin{center}{\sc #1} \end{center}}
\def\Address#1{\begin{center}{ \it #1} \end{center}}

\newcommand\pubblock{\rightline{\begin{tabular}{l} \pubnumber\\
         \pubdate  \end{tabular}}}
\newenvironment{Abstract}{\begin{quotation}  }{\end{quotation}}
\newenvironment{Presented}{\begin{quotation} \begin{center} 
             PRESENTED AT\end{center}\bigskip 
      \begin{center}\begin{large}}{\end{large}\end{center} \end{quotation}}
\def\Acknowledgments{\bigskip  \bigskip \begin{center} \begin{large}
             \bf ACKNOWLEDGMENTS \end{large}\end{center}}

\def\beq{\begin{equation}}
\def\eeq#1{\label{#1}\end{equation}}
\def\eeqn{\end{equation}}

\def\beqa{\begin{eqnarray}}
\def\eeqa#1{\label{#1}\end{eqnarray}}
\def\eeqan{\end{eqnarray}}

\let\bar=\overbar

\def\Dslash{\not{\hbox{\kern-4pt $D$}}}
\def\dslash{\not{\hbox{\kern-2pt $\del$}}}

\def\msb{{\bar{\ssstyle M \kern -1pt S}}}

\begin{document}
\begin{titlepage}
\pubblock

\vfill
\Title{Radiation Hard \& High Light Yield Scintillator Search for CMS Phase II Upgrade}
\vfill
\Author{Emrah Tiras\footnote{Corresponding author: emrah-tiras@uiowa.edu}} 
\center On behalf of the CMS Collaboration
\Address{\iowa}
\vfill
\begin{Abstract}
The CMS detector at the LHC requires a major upgrade to cope with the higher instantaneous luminosity and the elevated radiation levels. The active media of the forward backing hadron calorimeters is projected to be radiation-hard, high light yield scintillation materials or similar alternatives. In this context, we have studied various radiation-hard scintillating materials such as Polyethylene Terephthalate (PET), Polyethylene Naphthalate (PEN), High Efficiency Mirror (HEM) and quartz plates with various coatings. The quartz plates are pure \v{C}erenkov radiators and their radiation hardness has been confirmed. In order to increase the light output, we considered organic and inorganic coating materials such as p-Terphenyl (pTp), Anthracene and Gallium-doped Zinc Oxide (ZnO:Ga) that are applied as thin layers on the surface of the quartz plates. Here, we present the results of the related test beam activities, laboratory measurements and recent developments. 
\end{Abstract}
\vfill
\begin{Presented}
DPF 2015\\
The Meeting of the American Physical Society\\
Division of Particles and Fields\\
Ann Arbor, Michigan, August 4--8, 2015\\
\end{Presented}
\vfill
\end{titlepage}
\def\thefootnote{\fnsymbol{footnote}}
\setcounter{footnote}{0}

\section{Introduction}
LHC will reach up to ten times the design luminosity $(5-10\times 10^{34}cm^{-2}s^{-1})$ resulting in unprecedented radiation conditions in a collider experiment \cite{Collaboration2011}. The CMS Endcap calorimeters covering pseudorapidity between 1.6 and 3 will need to be replaced with a high granularity calorimeter and a backing hadron calorimeter during Phase II upgrades in order to maintain/improve the superior performance of CMS. 

The backing hadron calorimeter will have scintillator tiles as active media, and either direct coupling to the photodetectors to the tiles or utilization of WLS fibers. In this context, R\&D on radiation-hard, high light yield scintillators and radiation-hard WLS fibers will be crucial. 

Here, we attempt to identify the most suitable active media and readout options for this upgrade.

\section{Scintillator R\&D}

As the radiation-hard material, quartz is considered. Quartz plates are extremely radiation-hard \cite{Akgun2006}, but the signal generation based on measuring \v{C}erenkov light, hence is low level. The solution developed to overcome this issue is coating quartz plates with organic and inorganic scintillators such as para-Terphenyl (pTp), Anthracene (AN) and Gallium-doped Zinc Oxide (ZnO:Ga) in order to enhance the light yield.  

\begin{figure}[h]
\makebox[\textwidth]{%
\includegraphics[scale=0.4]{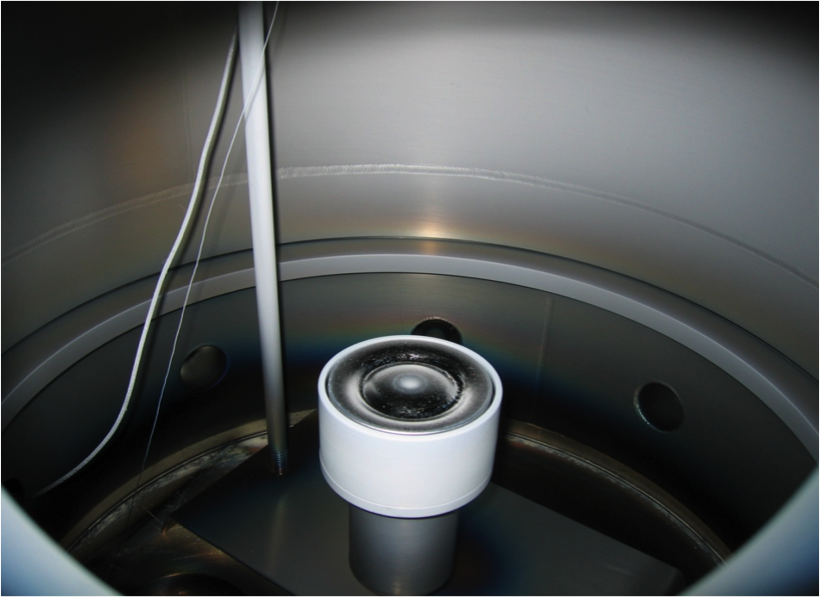}%
\\
\includegraphics[scale=0.4]{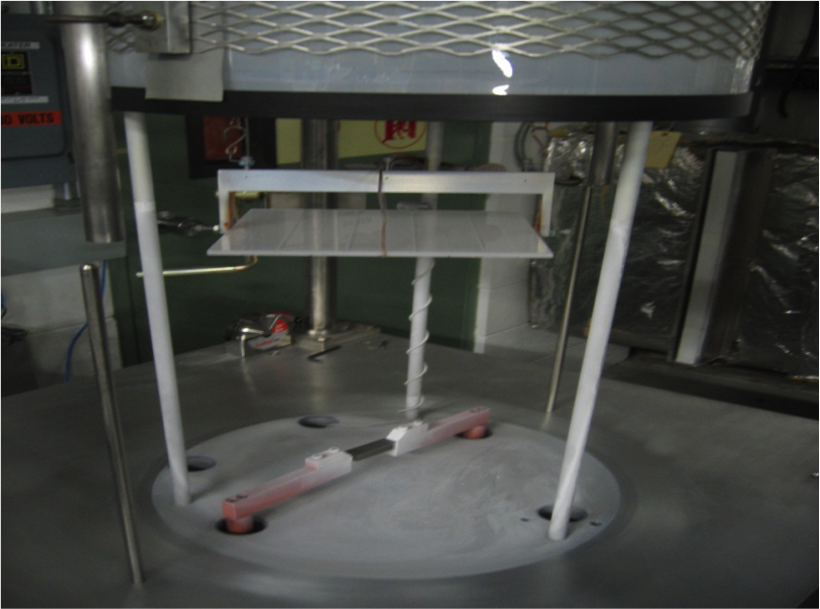}%
}%
\caption{pTp evaporation (left) and AN RF-sputtering (right) on quartz plates.}
\label{coatings}
\end{figure}

Organic scintillators, pTp and AN are aromatic hydrocarbons with three benzene rings $(C_{6}H_{6})$ formed. They exhibit blue fluorescence under UV. The evaporation technique is used for coating pTp and RF sputtering technique is used for coating AN on the quartz plates, Fig. \ref{coatings}. 

The radiation hardness of pTp was tested with proton beams at CERN and at the Indiana University Cyclotron Facility (IUCF). The light yield of pTp sample dropped to 84\% of the initial light yield after 20 MRad proton irradiation, and 80\% of the initial light yield after 40 MRad proton irradiation \cite{Bilki2010}.

Inorganic scintillators such as ZnO:Ga is also coated on quartz plates to increase the light yield. It has a short de-excitation time of 0.7 ns and very high luminous yield of 15k photon/MeV \cite{Derenzo2002}.

Also studied are intrinsically radiation-hard scintillators such as  Polyethylene Naphthalate (PEN), Polyethylene Terephthalate (PET) and High Efficiency Mirror (HEM). PEN and PET are bright and inexpensive plastic scintillators. 

\begin{figure}[h]
\centering
\includegraphics[scale=0.5]{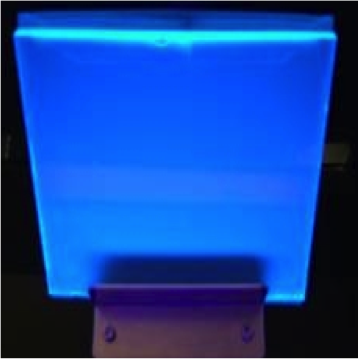}
\caption{The intrinsic blue scintillation of PEN}
\label{PENscintillation}
\end{figure}

PEN was created by the Japanese company Teijin Chemicals \cite{Teijin2015}. The company initially produced a sample in size of 5 mm x 35 mm x 35 mm and measured its light yield as 10,500 photons/MeV. PEN makes intrinsic blue scintillation, as can be seen in Fig. \ref{PENscintillation} with a peak emission spectrum of 425 nm \cite{nakamura2011evidence}.

PET is a common type of polyster and it is widely used to make plastic bottles and as a substrate in thin film solar cells. The emission spectrum of PET peaks at 385 nm \cite{Nakamura2013}. 

Another radiation-hard material is HEM, which is structurally a multilayer of polymer mirrors. We have made a stack of alternating slices of HEM sheet and quartz plates and tested the scintillating properties of the stack. 

\section{Test Beam Activities and Results}

\begin{figure}[h]
\makebox[\textwidth]{%
\includegraphics[scale=0.75]{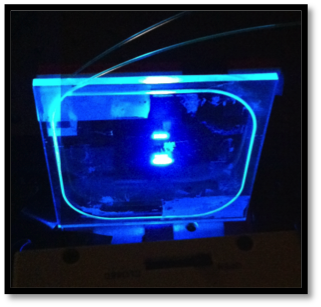}%
\\
\includegraphics[scale=0.204]{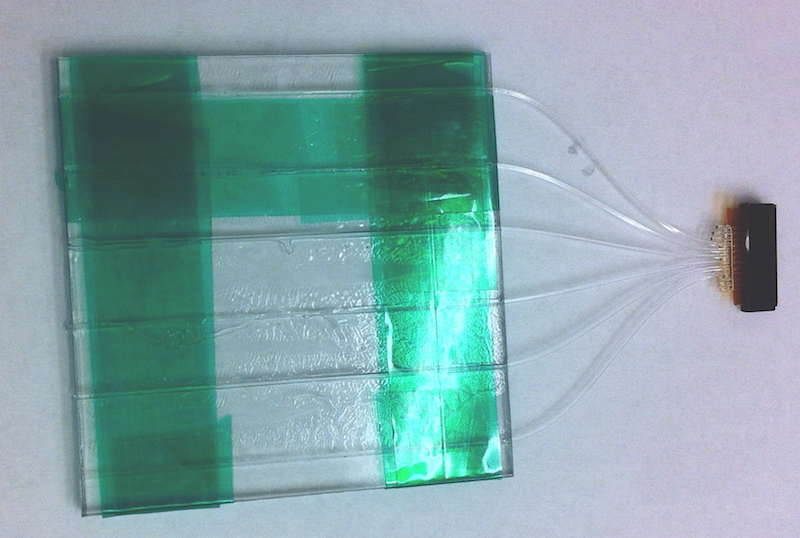}%
}%
\caption{Picture of sigma shape grooved tile with one WLS fiber (left) and a bar shape grooved tile with several WLS fibers (right).}
\label{fig:sigmabars}
\end{figure}

Various tiles were prepared and their timing characteristics, scintillation and transmission properties were studied at the University of Iowa Test Station. The tiles measured 10 cm$\times$10 cm and thickness of 1 mm and 2 mm. WLS fibers were coupled to the tiles with either sigma or bar shaped grooves. Figure \ref{fig:sigmabars} shows different tiles groove geometries. 

The test setup consisted of a light-tight box, a 334 nm wavelength UV laser, a Hamamatsu R7525 photomultiplier tube (PMT) \cite{R7525} and a Tektronix TDS 5034 digital oscilloscope. 

\begin{figure}[h]
\centering 
\includegraphics[scale=0.6]{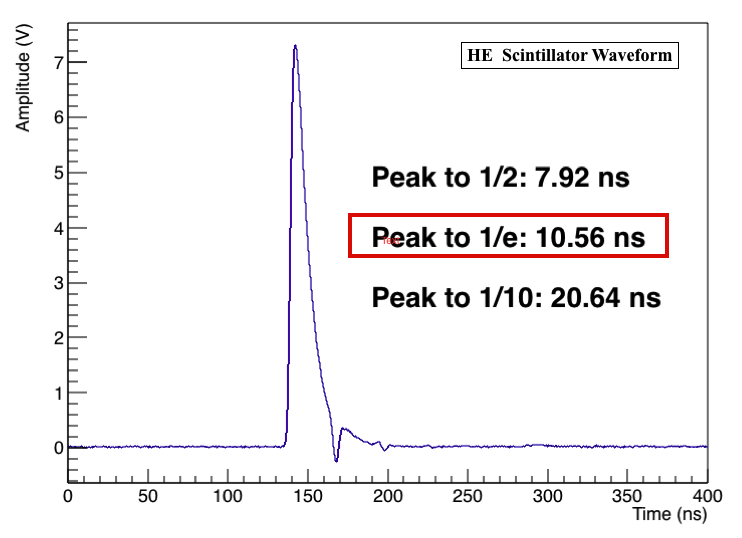}%
\caption{Signal timing of Kuraray SCSN-81 HE Tile.}
\label{fig:HETiming}
\end{figure}

\begin{figure}[h]
\makebox[\textwidth]{%
\includegraphics[scale=0.55]{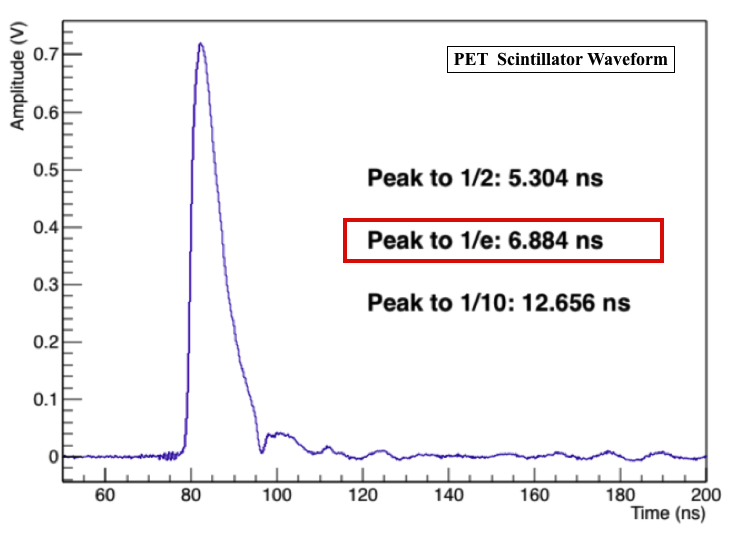}%
\\
\includegraphics[scale=0.58]{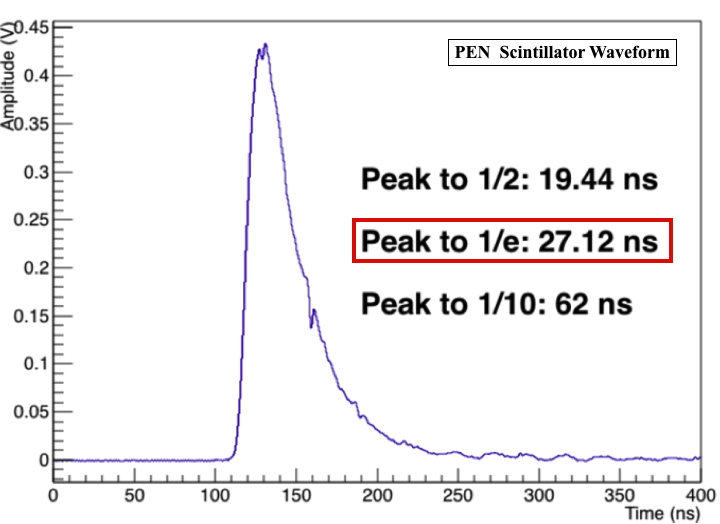}%
}%
\caption{Signal timing of PET (left) and PEN (right).}
\label{fig:PENPETTiming}
\end{figure}

Figure \ref{fig:HETiming} and \ref{fig:PENPETTiming} shows the timing characteristics of HE, PET and PEN. Signal timing was calculated, which is the time takes to fall from peak to half (1/2), from peak to 1/e and from peak to 1/10 of the signal. Peak to 1/e values for tiles, HE, PET and PEN are respectively 10.56 ns, 6.884 ns and 27.12 ns. PET has a much faster response than HE baseline tile, however PEN is slower.

\begin{figure}[!h]
\centering 
\includegraphics[scale=0.52]{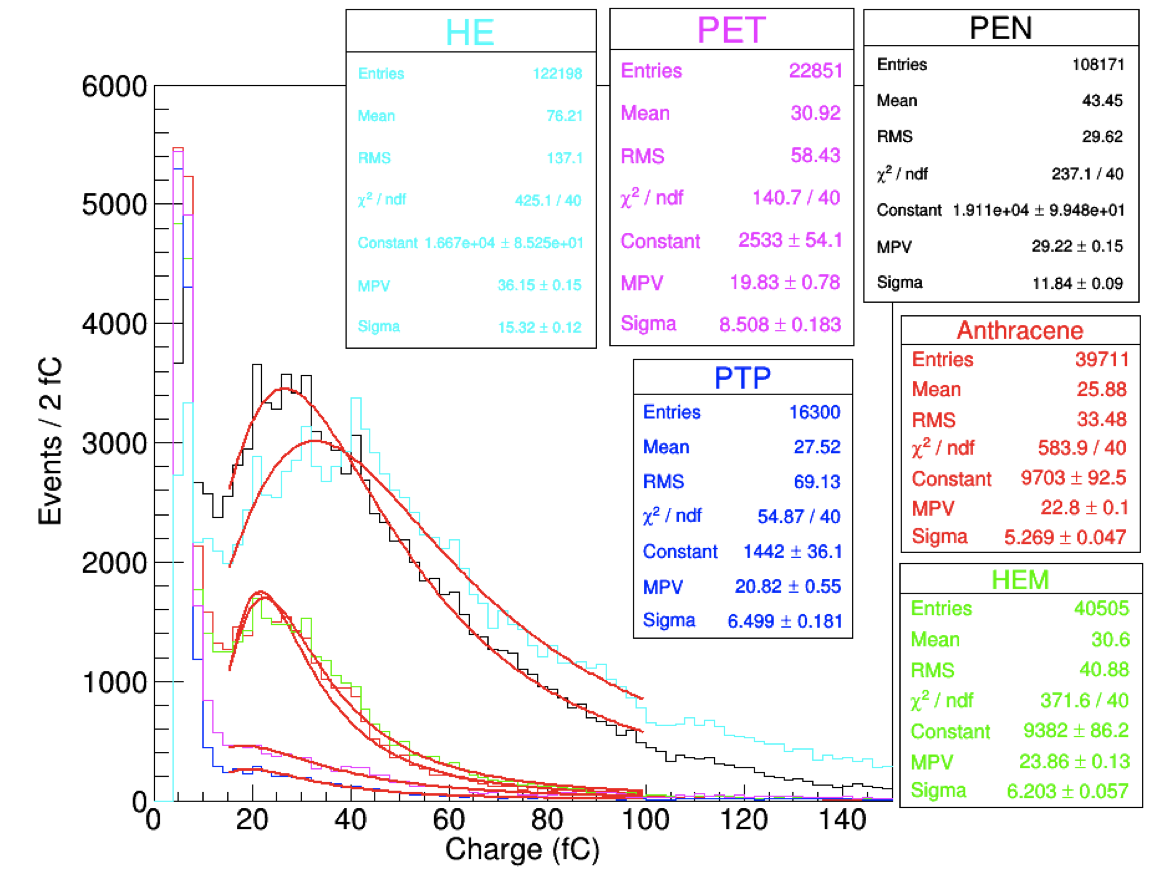}%
\caption{The MIP (muon) response of various tiles, tested at CERN H2 Test Beam Area.}
\label{fig:CERNResults}
\end{figure}

The assembled tiles were also tested at Fermilab Test Beam Facility (FTBF) and CERN H2 Test Beam Area for minimum ionizing particle (MIP) response. Figure \ref{fig:CERNResults} shows the MIP (muon) response of various tiles. 
For this test, the same PMT was used \cite{R7600PMT}. The most probable values (MPVs) of the Landau fits to the charge spectra above 15 fC for PEN, PET, HEM, coated quartz with pTp and AN are compared with the baseline HE tile. The measured MPV of HE is 36.15 fC and the other tiles have 29.22 fC, 19.83 fC, 23.86 fC, 20.82 fC and 22.8 fC respectively. PEN has the closest response compared to HE. The systematic effects associated with WLS fiber coupling has not been studied in detail.

\section{Conclusion}

Table \ref{tab:tiles} shows the MIP and timing response summary of HE, PEN and PET tiles. Although the light yield of PEN is much higher than PET, PET has a faster time response than PEN and SCSN-81 which is currently used as the active medium in the Hadron Endcap Calorimeters at CMS. A blended sample of PEN and PET was produced and tested by H. Nakamura, et al. and light yield of the blended substrate was measured 0.85 times that of PEN and much higher than that of PET \cite{Nakamura2013}. The blended sample is yet to be investigated for signal timing properties. 

\begin{table}[h]
\caption{Summary of HE, PEN, and PET comparison.}
\begin{center}
\begin{tabular}{l|ccc} 
Tiles &  SCSN-81 HE &  PEN &  
PET \\ \hline
 MIP Response (MPV, fC)  &   36.15     &     29.22      &     19.83  \\
 Timing Response (Peak to 1/e, ns) &  10.56     &     27.12      &  6.884 \\ \hline
\end{tabular}
\label{tab:tiles}
\end{center}
\end{table}

The other tiles, quartz with pTp and AN coatings and HEM, have also comparable light yields with HE. 

The R\&D is still underway in order to find the best radiation hard and high light yield materials. The extent of the R\&D is not limited to the future CMS upgrades but can find implementation areas in future collider detector experiments and in facilities where measurements in high radiation areas are crucial. 

\Acknowledgments
The author would like to thank Eileen Han at Fermi National Accelerator Laboratory for her assistance with coating quartz plates.


\end{document}